\documentclass[twocolumn,amsmath,floatfix,prl,aps]{revtex4-1}

\usepackage{color}
\usepackage{lipsum,amsmath}
\usepackage{pifont}   % Ding symbols
\usepackage{graphicx} % Include figure files
\usepackage{dcolumn}  % Align table columns on decimal point
\usepackage{bm}       % bold math
\usepackage{amsfonts} % Some more fonts
\usepackage{amssymb}  % More symbols
\usepackage{multirow} % Table functions
\usepackage{tikz}
\usepackage{siunitx}
\usepackage{placeins}
\usepackage{braket}
\usepackage{nicefrac}
\usepackage{bm}
\usepackage{physics}
\usepackage{grffile}

\usepackage{array,multirow,graphicx}
\usepackage[latin1]{inputenc}
\usepackage{caption}
\usepackage{scrextend}
\usepackage{hyperref}
\usepackage{float}
\usepackage{subcaption}
\usepackage{gensymb}
\usepackage[normalem]{ulem}

\newcommand{\ben}{\begin{equation}}
\newcommand{\een}{\end{equation}}

\def\bj{{\bf j}}
\def\bE{{\bf E}}
\def\bA{{\bf A}}

\begin{document}

%\title{Ab-initio calculations of the Giant Optical Intrinsic Spin Hall Effect in Bulk Transition Metals}
\title{The Giant Spin Hall Effect at Optical Frequencies}
\author{P. Elliott}
\affiliation{Max-Born-Institute for Non-linear Optics and Short Pulse Spectroscopy, Max-Born Strasse 2A, 12489 Berlin, Germany}
\author{S. Shallcross}
\affiliation{Max-Born-Institute for Non-linear Optics and Short Pulse Spectroscopy, Max-Born Strasse 2A, 12489 Berlin, Germany}
\author{S. Sharma}
\affiliation{Max-Born-Institute for Non-linear Optics and Short Pulse Spectroscopy, Max-Born Strasse 2A, 12489 Berlin, Germany}

\date{\today}
\begin{abstract}
We generalize the spin Hall angle to laser pulses of finite frequencies in the linear response regime and predict a giant optical spin Hall effect. Namely, for certain transition metal elements, at particular frequencies, the spin current can be a significant fraction of the charge current, and even exceed it for XUV frequencies. By maximizing spin current while minimizing the charge current, we thus minimize a major source of heating in spintronic devices. We employ {\it ab-initio} time-dependent density functional theory (TDDFT), and with real-time simulations calculate the conductivity and transverse spin conductivity for all $3$d, $4$d, and $5$d transition metals for frequencies up to $50$ eV. In the XUV frequency range we find values greater than $1$ for the spin Hall angle, indicating spin currents larger than the charge current can be generated.
%The spin-Hall effect (SHE) is a powerful technique to generate a spin current (perpendicular to a charge current) that may be used in spintronic devices such as SOT-MRAM. An important measure of the SHE is the spin Hall angle which measures the ratio between the charge current and the transverse spin current. We generalize the spin Hall angle for laser pulses of finite frequencies in the linear response regime and predict a giant optical spin Hall effect. Namely, for certain transition metal elements, at particular frequencies, then the spin current can be a significant fraction of the charge current, even more so than the giant SHE in the static dc limit. Thus, we maximize the spin current while minimizing the charge current, which in turn minimizes a major source of heating in spintronic devices. This is based on ab-initio time-dependent density functional theory (TDDFT) which can predict the intrinsic spin Hall effect. Using real-time simulations, we calculate the conductivity and transverse spin conductivity for all $3$d, $4$d, and $5$d transition metals for frequencies up to $50$ eV. In the XUV frequency range, we even find values greater than $1$ for the spin Hall angle, meaning spin currents larger than the charge current can be generated. Finally we show how the frequency range of the SHE can be controlled by combining elements into ordered alloys. 
\end{abstract}

\maketitle

%%%%%%%%%%%%%%%%%%%%%%%%%%%%%%%%%%%%%%%%%%
% INTRO
%%%%%%%%%%%%%%%%%%%%%%%%%%%%%%%%%%%%%%%%%%
%\section{Introduction}

In the field of spintronics, the electronic spin degree of freedom is utilized instead of, or in conjunction with, the charge degree of freedom to design more efficient technologies. While it is a relatively young research field, it has already had a significant impact on practical electronics thanks to the integration of the giant magnetoresistance\cite{BBFV88,BGSZ89} (GMR) effect into commercially available devices. Similarly, magnetoresistive RAM (MRAM) in which, unlike conventional RAM, bits are stored via the magnetic moment, is another highly promising area of spintronics. An emerging form of MRAM devices are based on the spin-orbit torque (SOT)\cite{MZMJ19}, where the magnetic moment is manipulated via a charge current flowing in the substrate, and not in the magnetic material itself\cite{MGGZ11}. This is possible when the substrate has strong spin-orbit coupling which can convert a charge current into a perpendicular spin current via the spin Hall effect. Thus, a key component of SOT-MRAM is the spin Hall effect (SHE).

In the ordinary Hall effect, discovered in $1879$, an external magnetic field perpendicular to a charge current generates an additional charge current perpendicular to both. The spin Hall effect\cite{DP71}, in contrast, produces a spin current perpendicular to both the charge current direction and the spin quantization axis in the absence of an external magnetic field. In $2004$ the accumulation of opposite spin polarizations on the two sides of a Hall bar was observed, confirming the existence of the spin Hall effect experimentally\cite{KMGA04,WKSJ05} and beginning the era of SHE physics. 

Typically the SHE has been studied\cite{OMSK92,SJKM03,GYN05,YF05,FBM10,GMCN08,FBM14,WWLY16,DNB17,TKD19,RPS19,HLLM13,TKNN08} and utilized in the static, zero-frequency, or dc, limit of electric fields. However, recently, the spin-Hall effect was studied for bilayers in the THz frequency regime\cite{GSBC17} and the
inverse spin-Hall effect was used to generate Terahertz pulses\cite{SJMH16} from current pulses themselves generated by femtosecond laser pulses. Extending the SHE into the domain of optical-XUV frequency electric fields, such as those in femtosecond laser pulses, will be of crucial importance in uniting the fields of spintronics and ultrafast magnetism\cite{B09}, and expected to drastically improve the speed and efficiency of spintronic devices. In this work, we calculate the spin-current tensor for non-collinear systems and extend the definition of the spin-Hall angle to the finite frequency domain. This allows us to predict a remarkable behavior for the frequency dependent spin current conductivities finding for several transition metals that the frequency dependent spin Hall angle greatly exceeds that found in the dc limit, and can even exceed unity, implying that spin currents of greater magnitude than the driving charge current can be generated. Such a situation represents the ideal physics for spintronics devices as the important spin current is maximised while the heat deposited due to Joule heating by the charge current is minimized.
 
The underlying physics of the spin Hall effect is typically due to the relativistic effect of spin-orbit coupling\cite{ERH07,SVWB15,RKB21}. In the intrinsic form of the SHE, spin-orbit coupling induces a spin texture (or Berry curvature) in the bands of the material, which then acts like a spin-dependent magnetic field deflecting moving electrons. In the extrinsic case, the additional spin-orbit interaction due to impurities cause difference scattering events that depend on the spin of an electron. In this work we will only consider the intrinsic effect.

Our method of choice is time-dependent density functional theory (TDDFT) which is an \emph{ab-initio} method for calculating electron dynamics\cite{RG84}. This method has had enormous success in both, explaining the underlying physics of ultrafast processes\cite{KDES15,FRBM14}, and also predicting new phenomena, such as OISTR\cite{EMDS16,DESG18}, which was subsequently experimentally verified\cite{SWGK20,CBEM19,WKSS20,SGOD19,HHDT20,GKGT21}. Thus, TDDFT is the ideal method to explore the behavior of the SHE at finite frequencies. For the choice of materials, we will explore all $3d$, $4d$, and $5d$ transition metals (except Mn due to the large number of atoms in the unit cell\cite{HH03}), as these have strong SOC and are known to display a strong SHE in the dc limit.

\begin{figure}[t!]
    \centering
    \includegraphics[width=0.8\columnwidth]{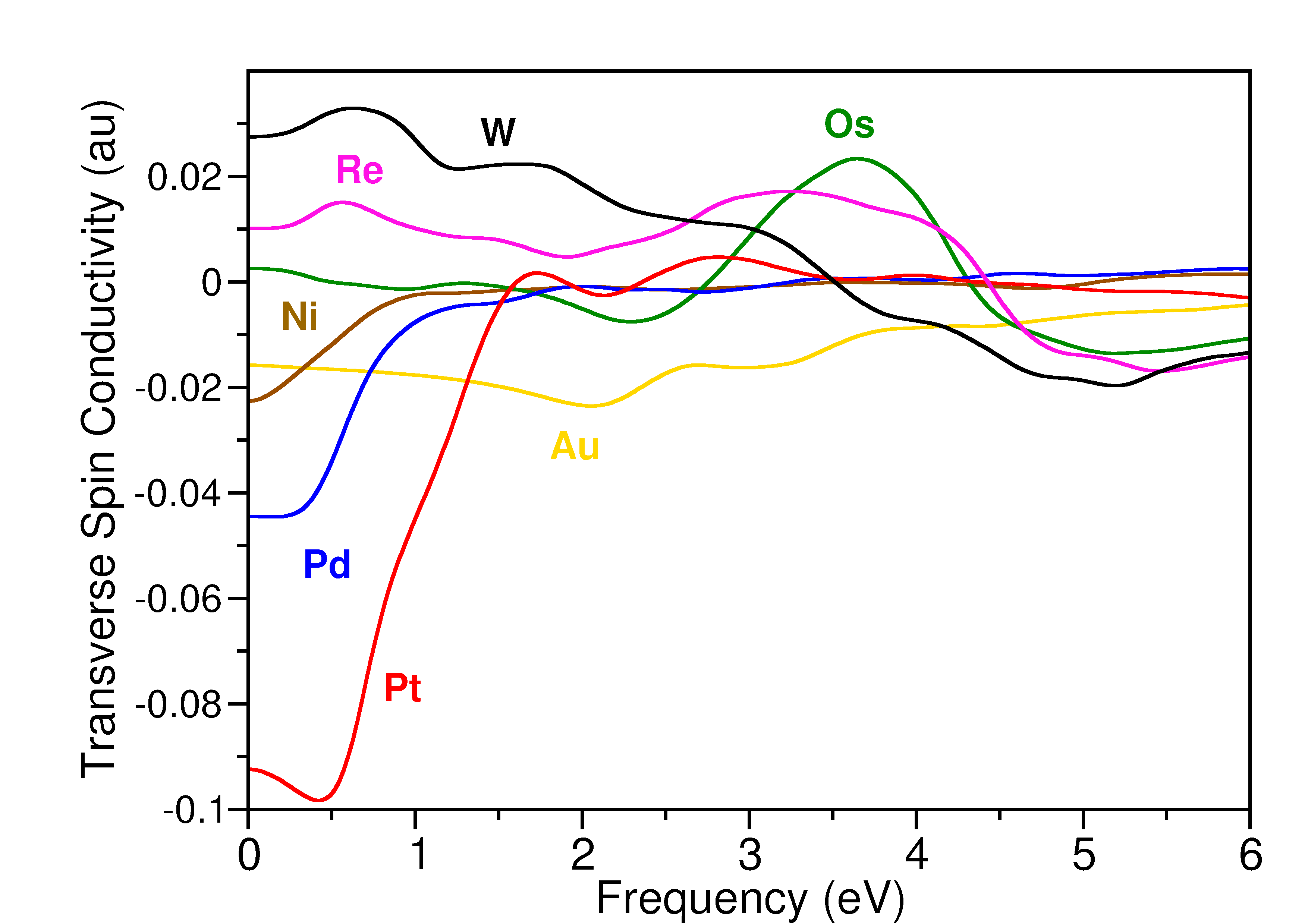}
    \caption{The frequency-dependent transverse spin conductivity for several transition metals calculated by real-time TDDFT in the linear response regime, shown for optical frequencies.}
    \label{f:sigmas}
\end{figure}

In TDDFT, the charge current and spin currents are calculated from the expectation value of the time-dependent Kohn-Sham wavefunction at each timestep. For more information on the propagation scheme, TDDFT\cite{RG84}, and the computational parameters\cite{HTB09,JLKV14,LRS15}, please see the supplementary material available at [URL will be inserted by publisher]. The charge current operator, $\hat{\bj}$ contains the usual paramagnetic and diamagnetic terms (as well as a small correction due to SOC). On the other hand the spin current operator (which measures spin flow in direction $a=x,y,z$ of spin along quantization axis $\mu=x,y,z$) is not uniquely defined\cite{VGW07}. A fact that follows from the absence of a continuity equation for the spin current and magnetization, due to non-conservation of the magnetic moment in the presence of SOC. The simplest choice is:

\ben
\hat{\mathbf{j}}^S_{a\mu} = \mathbf{\hat{j}}_a\boldsymbol{\hat{\sigma}}_\mu 
\een
which satisfies the continuity equation in the absence of SOC.

To calculate the charge conductivity tensor, ($\sigma_{ab}$), and the spin conductivity tensor, ($\sigma^S_{a\mu b}(\omega)$), in the linear response regime, a real time TDDFT calculation is performed using the ELK electronic structure code\cite{elk} for a weak perturbation and the resulting currents Fourier transformed to frequency space:

\begin{align}
\sigma_{ab}(\omega) &= \frac{\bj_a(\omega)}{\bE_b(\omega)} \\
\sigma^S_{a\mu b}(\omega) &= \frac{\bj^S_{a\mu}(\omega)}{\bE_b(\omega)}
\end{align}
where $\bE$ is a weak perturbing electric field, related to the vector potential by $\bE(t)=-1/c \ \textrm{d}\bA_{\rm ext}(t)/\textrm{d}t$ with $c$ the speed of light. 
%Typically a perturbation of $A_z = -c\kappa\theta(t)$, where $\theta(t)$ is the Heaviside step function and $\kappa$ is a small number, is used as allows us to determine the conductivities over a wide frequency range from a single calculation. 
The spin Hall effect  is measured by the term $\sigma^S_{xyz}(\omega)$, which we denote as the spin Hall conductivity or the transverse spin conductivity. It measures the spin current flowing in direction $x$ of the spin oriented along $y$ due to an applied electric field in the $z$ direction (which also drives a charge current in the $z$-direction).

\begin{figure}[t!]
    \centering
    \includegraphics[width=0.8\columnwidth]{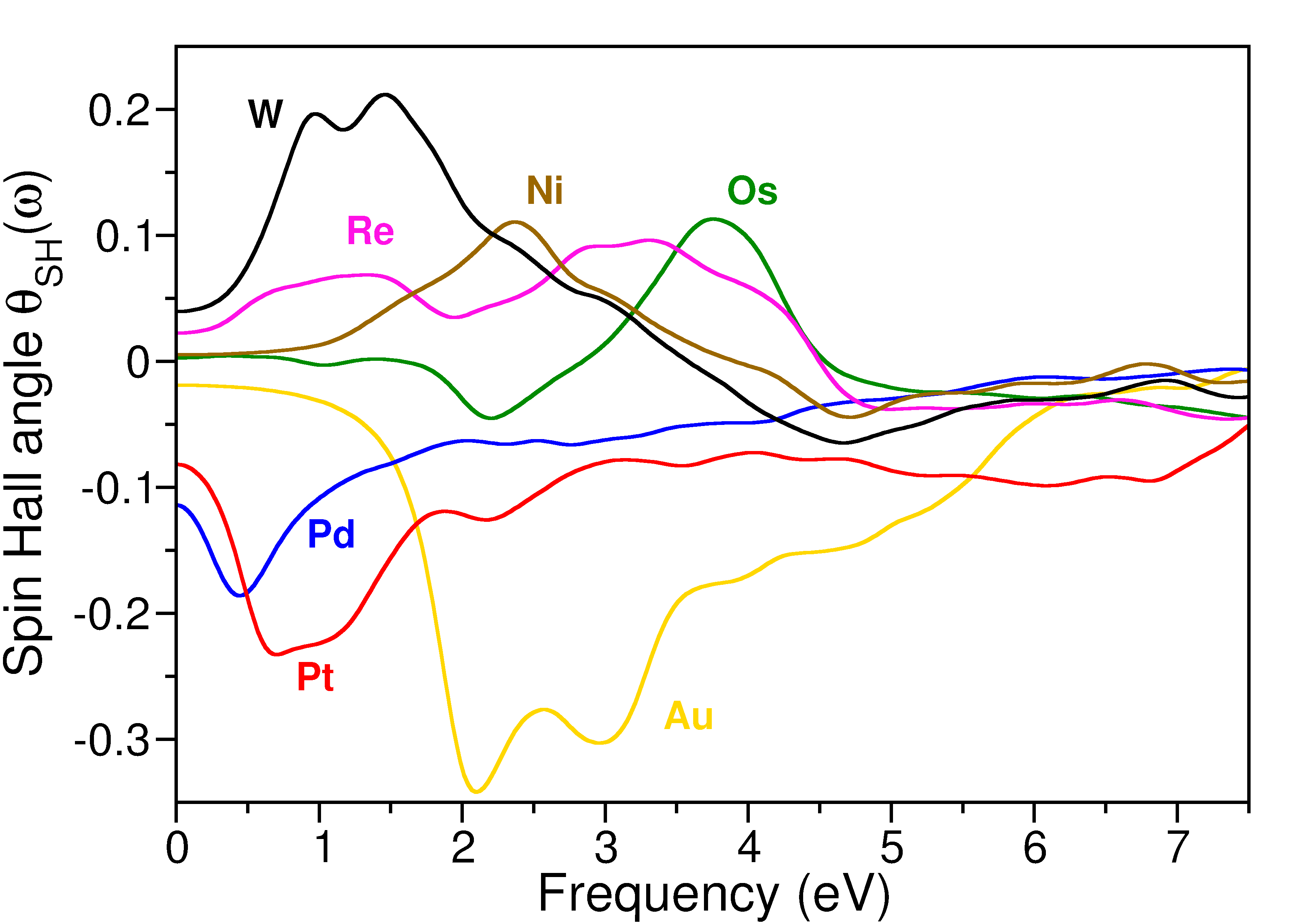}
    \caption{The spin Hall angle, defined as the ratio of the transverse spin conductivity and the charge conductivity, or equivalently the Fourier transformed perpendicular spin current divided by the charge current, for several transition metals. The giant optical spin Hall effect can be seen for the large values of the spin Hall angle at optical frequencies.}
    \label{f:thetas}
\end{figure}

%%%%%%%%%%%%%%%%%%%%%%%%%%%%%%%%%%%%%%%%%%
% RESULTS
%%%%%%%%%%%%%%%%%%%%%%%%%%%%%%%%%%%%%%%%%%

In Fig.~\ref{f:sigmas}, we show the transverse spin conductivity for those transition metals with the largest SHE (for others see SI). As can be seen, there is significant frequency dependence. 
At low frequencies (from $0$ to $1$ eV), W, Pd, and Pt have the largest spin Hall conductivities, unsurprising given these elements display a giant SHE in the dc limit\cite{MNOW11,SOIG16,PLLT12,HX15,QHGC18}. It is interesting to note that the spin conductivity is finite at $\omega=0$, in contrast to the charge current where electron-phonon scattering is required to prevent the current diverging in this limit. This highlights the difference between the physics of charge current and spin currents. Moving to higher frequencies, we see that other materials such as Au, Re, and Os show the best spin current response to an electric field, although W remains strong for a wide range of frequencies (even after it switches sign). For the optical range of $1.5-3.0$~eV, Au and W exhibit the largest spin currents. 
%Predicting the SHE using \emph{ab-initio} methods has previously been performed in the dc limit\cite{OMSK92,SJKM03,GYN05,FBM10,GMCN08,FBM14,WWLY16,DNB17,TKD19,RPS19,HLLM13}, although it has also been calculated at finite frequencies for Pt and W \cite{SMBF17}. In these cases, the Kubo independent-particle formalism was used, which can be considered an approximation to the, in principle exact, TDDFT result.
We note that the Kubo-Greenwood independent-particle formalism results for the W and Pt spin conductivities calculated in Ref.~\cite{SMBF17} agree with the TDDFT results of Fig.~\ref{f:sigmas}, indicating that exchange-correlation corrections are small, at least in this case. 

In the dc case, the ratio between the spin current and the charge current is known as the spin Hall angle (SHA) and is commonly used to characterize the SHE. It may be generalized\cite{GSBC17} to finite frequencies:

\ben
\theta^{\rm SH}(\omega) = \frac{\bj^S_{xy}(\omega)}{\bj_z(\omega)} = \frac{\sigma^S_{xyz}(\omega)}{\sigma_{zz}(\omega)}
\een

\begin{figure}[ht]
    \centering
    \includegraphics[trim=0 240 0 50,clip,width=\columnwidth]{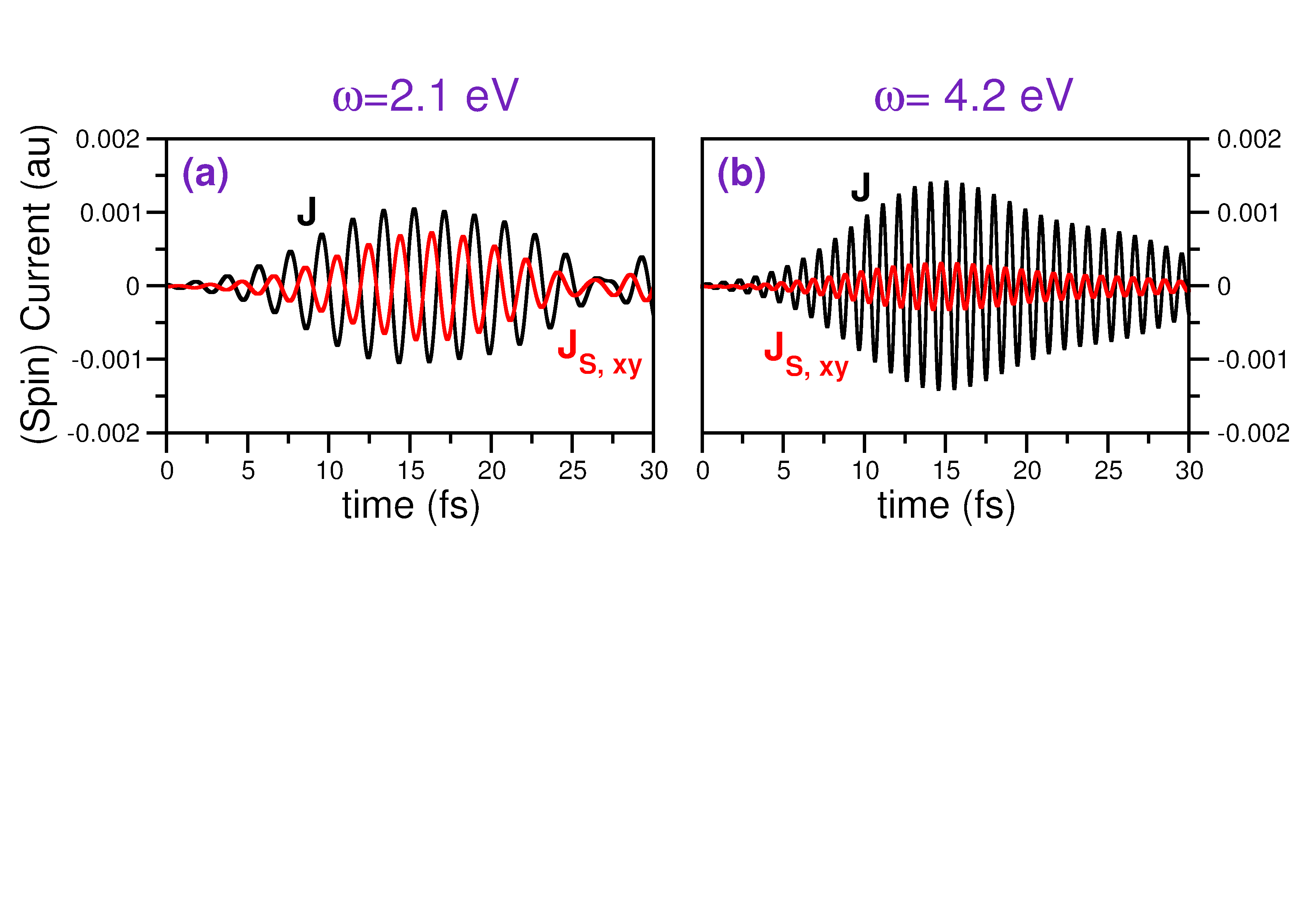}
    \caption{The charge and transverse spin currents in Au induced by laser pulses of frequency (a) $2.1$~eV and (b) $4.2$~eV. This demonstrates the giant optical spin Hall effect in Au for laser pulses with frequency $2.1$~eV.}
    \label{f:Au_Curr}    
\end{figure}

In Fig. \ref{f:thetas}, we plot the transition metal elements with the largest SHAs (for other metals see SI). Immediately we see a much richer structure than the spin conductivity alone, and find the interesting result 
%The SHA at a particular frequency depends on a number of factors. The charge current will depend on the states available for excitation at that frequency, the transition dipole moment between them, and the current carried by each state. This is summed over all occupied states in the entire first Brillouin zone (1BZ), and thus it is often difficult to assign a reason why the current response is large or small at a particular frequency. The spin-current adds the spin-texture/Berry curvature of the bands on top of these factors> Thus the SHA will have a complex dependence on the laser frequency.
%Interestingly, in Fig.~\ref{f:thetas} we see 
that the SHA at certain frequencies is very large and can even exceed the values of W and Pt at zero frequency. Indeed, even W  and Pt themselves have higher values at finite frequency than at zero frequency. We name this phenomenon as the giant optical spin Hall angle (this term has also been used in the context of a polariton spin Hall effect\cite{KMG05}, however it is the natural name for the effect described in this work). An enhancement of the dynamical spin Hall angle was also observed in Ref. \onlinecite{GSBC17} in the THZ frequency regime due to changes in the band structure induced by ferromagnetic layers.

To demonstrate how this giant optical spin Hall would behave in an experimental setup with a finite duration pulse, we simulate the response of Au to two different frequency pulses. From Fig.~\ref{f:thetas}, we see that Au has the largest SHA at $2.1$~eV and a value roughly half at $4.2$~eV. Thus in Fig.~\ref{f:Au_Curr}, we compare the charge current and transverse spin current induced by lasers oscillating at these frequencies. Both pulses have a duration full width at half maximum (FWHM) of $12.1$~fs and a peak intensity of $10^9$ W/cm$^2$.

In Fig.~\ref{f:Au_Curr} (a), we see that at $\omega=2.1$~eV, the charge and spin currents are of similar magnitude, while in Fig.~\ref{f:Au_Curr} (b), the charge current is much larger than the spin current. 
This is exactly as expected from Fig.~\ref{f:thetas}, which says that the ratio of the spin current to the charge current is larger at $\omega=2.1$~eV than at $\omega=4.2$~eV. In this case the amplitude of the charge current at both frequencies is similar, meaning the spin current at $\omega=2.1$~eV is much larger. 
We can also see that the charge and spin currents are $180^{\circ}$ out of phase with each other, as the SHA is negative. In both cases, the current oscillations will continue after the laser pulse, however decoherence and relaxation to the ground-state will damp these out on the $100$~fs to $1$~ps timescale. 

\begin{figure}[ht]
    \centering
    \includegraphics[trim=0 270 0 0,clip,width=\columnwidth]{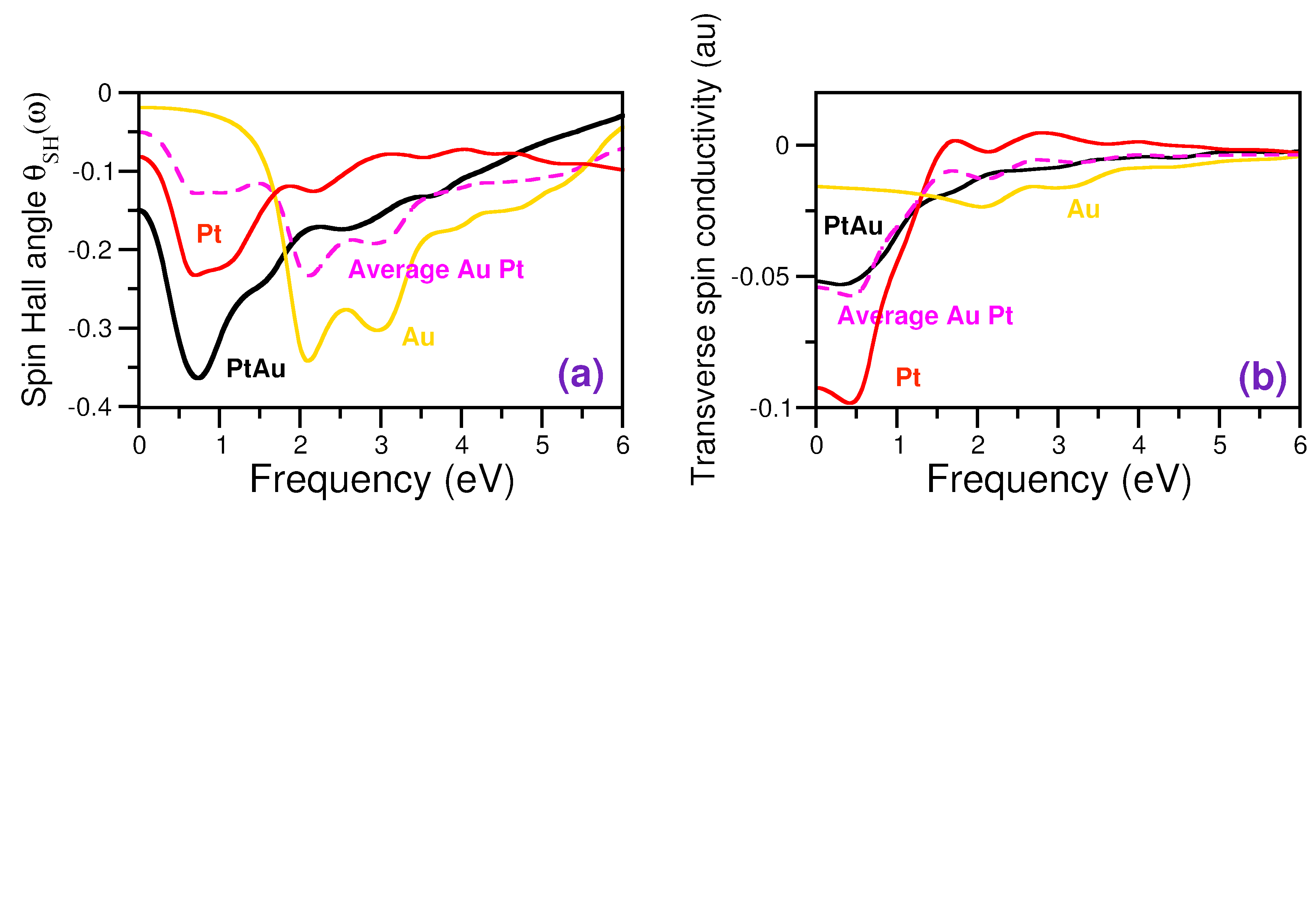}
%    \fbox{\includegraphics[trim=0 0 40 30,clip,width=0.45\columnwidth]{PtAu.png}}
%    \fbox{\includegraphics[trim=0 0 90 30,clip,width=0.4\columnwidth]{SHA-PtAU.png}}
    \caption{Tuning the spin Hall properties via ordered alloying of Pt and Au showing the frequency dependence of the (a) spin Hall angle and (b) the transverse spin Hall conductivity. For comparison, we show the elemental values for Au and Pt, as well as their their average, in each case.}
    \label{f:PtAu}    
\end{figure}

The giant SHA can be interpreted as the appropriate frequency to maximize the spin current for a given charge current amplitude, or equivalently minimize the charge current for a desired amplitude of spin current. 
This has great significance as the smaller the charge current, the less energy deposited into the system due to Joule heating. As dissipation of heat is one of the limiting factors currently hindering technological development, the giant SHA can be utilized in the quest for more efficient devices. One such device where the giant SHA can be exploited would be ultrafast SOT devices, where a magnetic moment is controlled via spin-currents generated from ultrafast laser pulses. In a SOT geometry, a laser incident on a SHE layer could create an oscillating spin-current that penetrates the magnetic layer. The induced dynamics of the magnetic moment due to such a spin current is an open question, which will be studied in future work.  

A key quality required for efficient technological transfer is the ability to tailor the material for desired properties. For example, whether transition metal elements could be combined to enhance or extend the frequency range of the SHE. From Fig.~\ref{f:sigmas}, we see that while both Pt and Au have large spin Hall conductivities, they occur in different frequency ranges: $0$ eV to $2$~eV for Pt, and $2$~eV to $4$~eV for Au. In Fig.~\ref{f:PtAu}(b) we show that mixing Pt and Au together in an ordered alloy (L$1_0$ geometry with the Au lattice spacing) produces a material with a large spin Hall conductivity over the entire NIR, visible, and VUV range between $0$ eV to $4$eV. In Fig.~\ref{f:PtAu}(b), we also see that the combined PtAu spin Hall conductivity is simply the average of the two independent Au and Pt values. However this will not be the case for the spin Hall angle. In Fig.~\ref{f:PtAu}(a), we see the remarkable result that the giant spin Hall angle seen in Pt at $0.75$ eV can be further enhanced in the PtAu alloy. This is evidently due to the changes in the charge conductivity caused by the altered band structure of the alloy. This new value of this giant spin Hall angle exceeds all those seen in the elemental cases of Fig.~\ref{f:thetas}, demonstrating how simple alloying can be used to tailor and optimize the SHE.

%We anticipate that for more systems with a more complicated mixing where the electronic structure is significantly altered, then novel features will appear in the conductivity. 

% enhances 
% spin conductivity roughtly average but charge conductivity changes as bands rearrange??

\begin{figure}[ht]
    \centering
    \includegraphics[width=0.8\columnwidth]{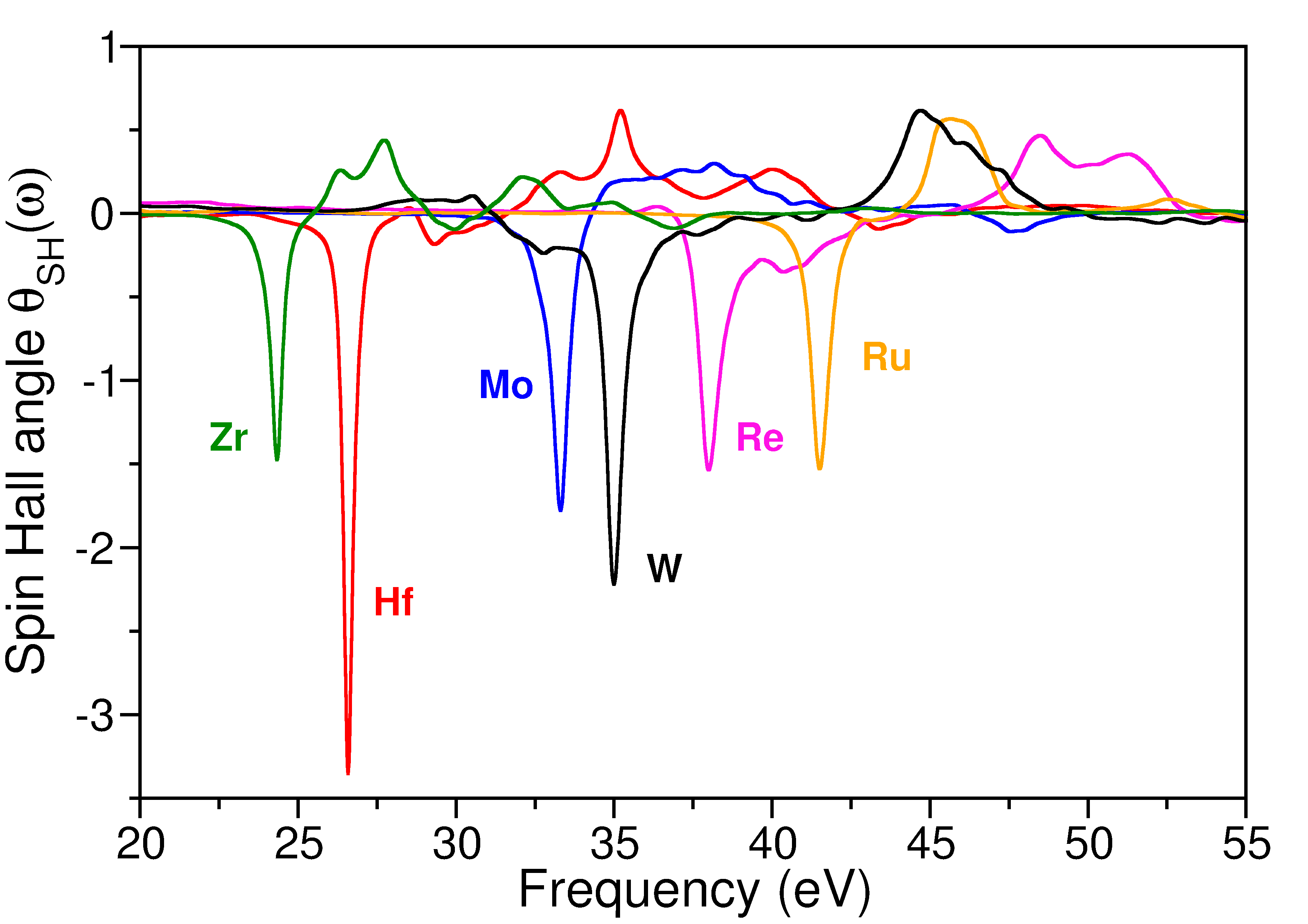}
    \caption{The spin Hall angle, as in Fig.~\ref{f:thetas}, extended to the XUV frequency range, where values several times larger than unity can be seen.}
    \label{f:thetaXUV}    
\end{figure}

At near infra-red (NIR) and optical frequencies, we saw in Fig.~\ref{f:thetas} that large spin Hall angles are possible at finite frequencies. The largest of these was Au with a value of $-0.4$ or a spin current magnitude of about $40\%$ of the charge current. We now show that in the x-ray ultraviolet (XUV) range, spin-currents several times larger than the charge current are possible.

In Fig.~\ref{f:thetaXUV}, we extend the spin Hall conductivities of Fig.~\ref{f:thetas} into the XUV frequency range up to $55$~eV. The dominant transition in this frequency range is from the $3p/4p/5p$ orbitals to the unoccupied
$3d/4d/5d$ orbitals at the Fermi level, otherwise known as the M$_{2,3}$, N$_{2,3}$, and O$_{2,3}$ absorption edges in XAS.
As with the optical frequency range, the charge and spin currents are determined by the interplay between the spin texture/Berry curvature of the occupied and unoccupied states in the excitation, with the added complication of strong spin orbit splitting of the p states. However, as we see in Fig. \ref{f:thetaXUV} this combination can create scenarios where a spin current is generated that is even larger than the charge current, i.e. a SHA above/below $\pm1$.

Only the materials with the largest XUV SHA are shown in Fig.~\ref{f:thetaXUV}. Curiously these materials are a mix of $4$d and 5$d$ elements, in contrast to the results presented in Figs.~\ref{f:sigmas} and \ref{f:thetas}, which are mostly $5$d elements. The exceptions being W and Re, which appear in all $3$ figures.

\begin{figure}[t]
    \centering
    \includegraphics[width=0.8\columnwidth]{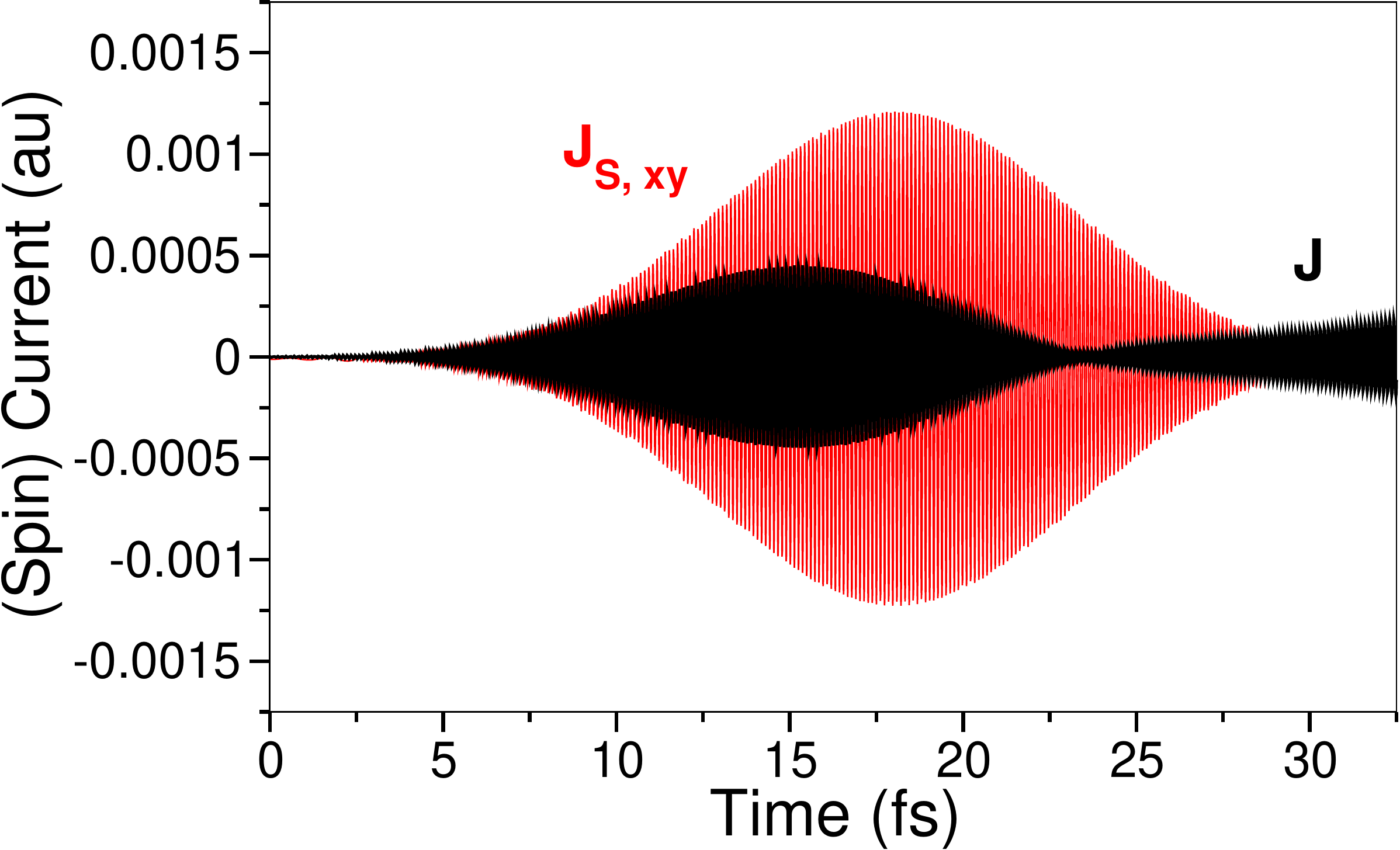}
    \caption{The charge and transverse spin currents in W induced by a laser with frequency $35$~eV. The amplitude of the spin current is several times larger than the charge current.}
    \label{f:W_Curr}        
\end{figure}

We again examine the behavior of the currents when using a finite duration pulse with a frequency centered at one of the SHA peaks. %To confirm this observed SHA in W, we again create a finite duration pulse at the frequency of the strong peak. 
In Fig.~\ref{f:W_Curr} we show the charge current and the perpendicular spin current induced by a pulse with frequency $\omega=35$~eV, a FWHM of $12.1$~fs, and a peak intensity of $10^{10}$~W/cm$^2$. Additionally, for such a high frequency pulse, the timestep in our time propagation algorithm must be decreased to $\Delta t=0.05$~a.u., to ensure accuracy and stability. As expected from Fig.~\ref{f:thetaXUV}, the spin-current is more than double the charge current, thus demonstrating a giant XUV spin Hall angle. In the supplemental material, we show the charge and spin conductivities for each element. This giant SHA is due to the charge current becoming small for a perturbation at this frequency while the spin-current response is similar to that in the optical range.

%%%%%%%%%%%%%%%%%%%%%%%%%%%%%%%%%%%%%%%%%%
% conclusions
%%%%%%%%%%%%%%%%%%%%%%%%%%%%%%%%%%%%%%%%%%
In conclusion we have calculated the charge and transverse spin-Hall conductivities in the linear response regime using real-time TDDFT for laser pulses in the NIR, optical, and XUV frequency ranges for almost all transition metal elements. We find the element with the highest spin response strongly depends on this frequency. For example, at optical frequencies, it is Au and W which show the strongest SHE unlike the dc limit where Pt is the strongest.

We have generalized the concept of the spin Hall angle to finite frequency using the Fourier transform of the induced charge current and perpendicular spin current. We found a giant intrinsic spin Hall angle exists for certain materials with laser pulses at particular frequencies, and have demonstrated this effect in Au where the ratio between the charge current and spin current can be maximized by choosing the pulse frequency. This minimizes a major source of heating in spintronic devices, namely Joule heating demonstrating the promise of combining the Spin Hall physics with ultrafast laser pulses. In future work, we will calculate the induced magnetization dynamics in SOT devices due to these ultrafast spin-currents. We also observed that for XUV frequencies, in the range $20-50$ eV, SHE spin currents could be generated that even by several times the charge current in amplitude. Finally we shown that by designing multi-component alloys we could enhance and extend the range of frequencies showing a significant SHE beyond those found in the constituent components. In future work we plan to study more complicated structures, in particular how the spin Hall conductivity depends on the interface.

\section{Acknowledgments}

PE thanks DFG for funding through project 438494688. Shallcross would like to thank DFG for funding through SPP 1840 QUTIF Grant No. SH 498/3-1, and Sharma thanks DFG for funding through project-ID 328545488 TRR227 (project A04).

\end{document}